# UNDERSTANDING ENTANGLEMENT AND RESOLVING THE MEASUREMENT PROBLEM

Art Hobson
Department of Physics, University of Arkansas,
Fayetteville, AR, 72701, USA
email: ahobson@uark.edu

ABSTRACT:   We summarize a recently proposed resolution of the quantum measurement problem.  It stems from an insight into entanglement demonstrated in a 1991 experiment involving photon momenta.  This experiment shows that, when two superposed quantum systems *A* and *B* are entangled, the resulting "pre-measurement state" is not a paradoxical macroscopic superposition of *compound states* of the two subsystems; for example, Schrodinger's cat is not "smeared" between dead and alive.  It is instead a non-local superposition of *correlations between states* of the subsystems.  In Schrodinger's example, an undecayed nucleus is correlated with a live cat, AND a decayed nucleus is correlated with a dead cat, where "AND" indicates the superposition.  This is exactly what we want.  We have misinterpreted "dyads" |A> |B> where "A" and "B" are subsystems of a composite system AB.  A> |B> does not mean states |A> and |B> both *exist*.  It means instead |B> exists if and only |A> exists, i.e. |A> and |B> are *correlated*.  It is a fact of nature that such correlations are *nonlocally coherent*:  The degree of correlation between A and B depends on the nonlocal phase angle between the arbitrarily distant subsystems.  Such coherent correlations are central to the nonlocal collapse of wave functions.

## INTRODUCTION

This analysis, which first appeared in more rudimentary form in [Hobson 2024] Chapters 9 and 10, concludes that the measurement problem arises from a misunderstanding of entangled states. We review two 1991 investigations of momentum-entangled photons. These experiments reveal that entangled states have been mis-interpreted. We show that a proper understanding of entanglement resolves the measurement problem.

This paper re-states the measurement problem, describes the misunderstanding of entanglement as revealed by experiment, and presents the proper understanding of entanglement, namely that entangled states are coherent (i.e., phase-dependent) superpositions of correlations between quantum states. Thus, the controversial "pre-measurement state" of a quantum object and its detector is not an absurd superposition of detector states; it is instead a perfectly plausible superposition of correlations between detector states and quantum states of the object.

## THE MEASUREMENT PROBLEM

We begin with a standard definition of the quantum measurement problem [MYRVOLD 2022]. Consider a quantum system *A* having (for simplicity) a two-dimensional Hilbert space spanned by orthonormal states |A1> and |A2>, and let **O** be the observable whose eigenstates are |A1> and |A2>. A "detector" D of **O** must contain a quantum component having three quantum states: $|D_{ready,}>$ represents a state in which D is poised to detect whether *A* is in state |A1> or |A2>, and |Di> (i=1 or 2) represents macroscopic registration that *A* is in the state |Ai>. D must also have a component that amplifies the microscopically detected outcome to irreversibly register that outcome, perhaps by

creating a visible mark or an audible click. Thus, *D* is a macroscopic object with a quantum component.

As an example, *A* might be a single electron passing through a double-slit setup with a viewing screen, with "which-slit detectors" $D_1$ and $D_2$ present at both slits. The states |Di> (i=1 or 2) then represent the "clicked" state of the first or second detector.

Suppose that, before measurement, *A* is prepared in an eigenstate |Ai> (i=1 or 2). A minimally disturbing measurement is then represented by

$$|Ai\rangle \, |D\,ready\rangle \implies |Ai\rangle \, |Di\rangle \quad (i = 1, 2) \tag{1}$$

where |Ai> represents the premeasurement eigenstate, the arrow represents the measurement process, and the right-hand side represents the post-measurement state. Note that the same state |Ai> appears on both sides of (1), i.e. we assume that, when *A* is prepared in an eigenstate of **O,** measurement of **O** does not disturb that eigenstate. This is an idealization.

Now suppose *A* is prepared in a 50-50 superposition of these eigenstates:

$$|\Psi_A\rangle = (|A1\rangle + |A2\rangle)/\sqrt{2}. \tag{2}$$

It follows from the linearity of the time evolution that a "which state" measurement of *A* is then represented by

$$\frac{(|A1\rangle + |A2\rangle)}{\sqrt{2}} |D\,ready\rangle \implies |\Psi_{AD}\rangle \tag{3}$$

where |$\Psi_{AD}$> is defined as

$$|\Psi_{AD}\rangle = \frac{|A1\rangle\,|D1\rangle + |A2\rangle\,|D2\rangle}{\sqrt{2}}. \tag{4}$$

This enigmatic entangled state crops up in nearly every analysis of the measurement problem.   We will call it the "pre-measurement state." [SCHLOSSHAUER 2007]

|$\Psi_{AD}$> appears to be a macroscopic quantum superposition of both detector states |$D_i$>.  If so, |$\Psi_{AD}$> would be a state in which both detectors clicked, indicating that the single electron was detected at both slits.  Such a superposition is not observed and would be absurd.

According to [MYRVOLD 2022], the measurement problem is "The problem of what to make of the pre-measurement state" (4). This paper concurs, and adopts Myrvold's definition.

According to conventional wisdom, the "plus" sign in (4) represents a superposition of two states of the compound system *AD*.  Those two states are represented by the dyads |$A_i$> |$D_i$> (i = 1, 2). For example, in the electron 2-slit experiment with detectors at both slits, the conventional interpretation of (4) would be the following:

In a single trial, the electron was detected at the first slit
AND the electron was detected at the second slit.             (5)

where "AND" represents the superposition. This describes an absurd superposition of two macroscopic outcomes.  What's wrong?

## A PROPOSED SOLUTION

Two quantum optics experiments published nearly simultaneously in 1991 furnish evidence that the conventional interpretation (5) of the measurement state (4) is incorrect [RARITY 1991, OU 1991].  We will call these experiments the "RTO experiments," honoring the two authors of the first report and the lead author of the second report.  RTO investigated entangled momentum states of two photons.  If we label the first photon "A" and the second photon "B," this entangled state was

$$|\Psi_{AB}\rangle = \frac{|A1\rangle |B1\rangle + |A2\rangle |B2\rangle}{\sqrt{2}} \ . \tag{6}$$

Equation (6) is isomorphic to (4) with the important distinction that both subsystems are now microscopic.  As we shall soon see, the entangled state (6) exhibits non-locality.

2022 Physics Nobel laureate Alain Aspect has remarked [PHILLIPS 2023] that the RTO experiments were unique because they were and are the only investigation of entangled *mechanical* degrees of freedom, namely linear photon momenta.  All other entanglement experiments, including Aspect's prize-winning experiment, investigated photon polarization states.

[HOBSON 2024] Chapter 9 reviews the RTO experiments.  In these experiments, two photons A and B were entangled by down-conversion of a single high-frequency photon.  A then moved along two beam-split paths, and B moved along two other distant beam-split paths.  Variable phase shifters $\phi_A$ and $\phi_B$ were inserted into one of A's two paths and one of B's two paths, respecitvely.   We label these phase shifts $\phi_A$ and $\phi_B$ (see Figure 1, which does not show the beam splitters).

Remarkably, *neither photon Interfered with itself as a function of its own phase $\phi_A$ or $\phi_B$*.  That is, *individual photons were incoherent and not phase-dependent or "smeared"*: Regardless of the phase settings $\phi_A$ and $\phi_B$,

$$P(A1) = P(A2) = P(B1) = P(B2) = 0.5 \tag{7}$$

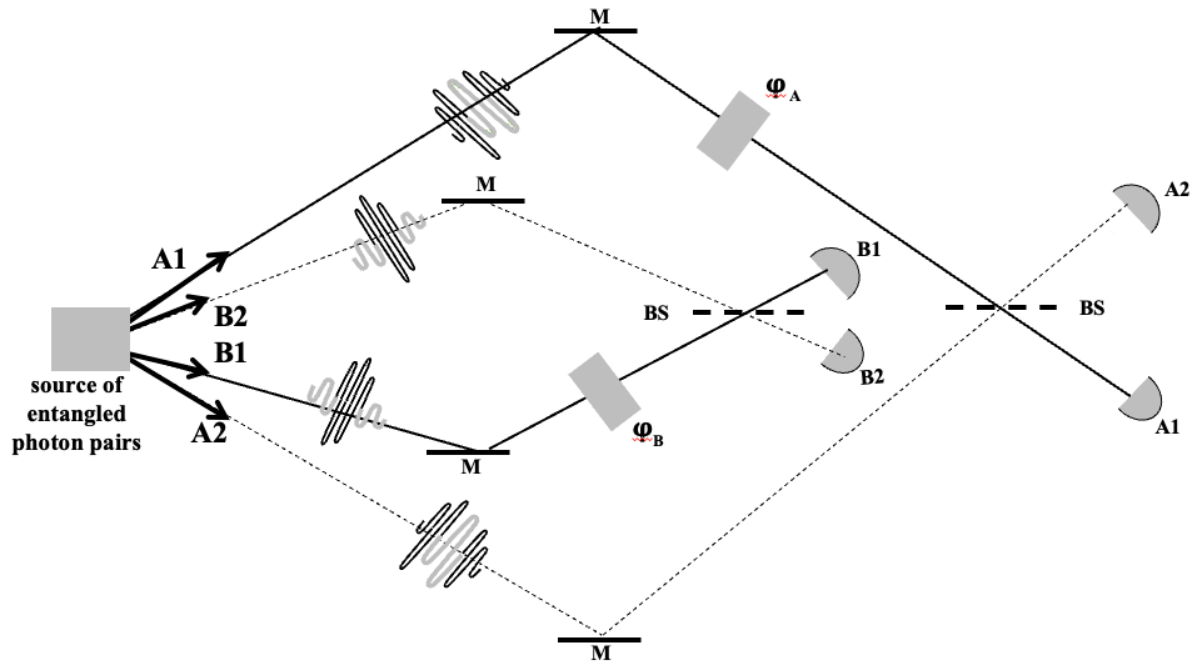

Figure 1. Layout of the RTO experiments. One photon emerges from the source along the beam-split paths A1 and A2; the other photon emerges along the beam-split paths B1 and B2. The two photons form a single entangled "biphoton."

This incoherence must be attributed to the entanglement.

However, *the coherence had not vanished*. RTO found that the expected coherence of each photon had instead *shifted*: *A* and *B* now interfered *with each other* (rather than with themselves) as a function of the *difference* $\phi_A - \phi_B$ of the two phase angles: When the experimenters shifted this "nonlocal phase difference" to various angles between 0 and 180 degrees, *the correlation between the two photons varied as shown in Figure 2.*

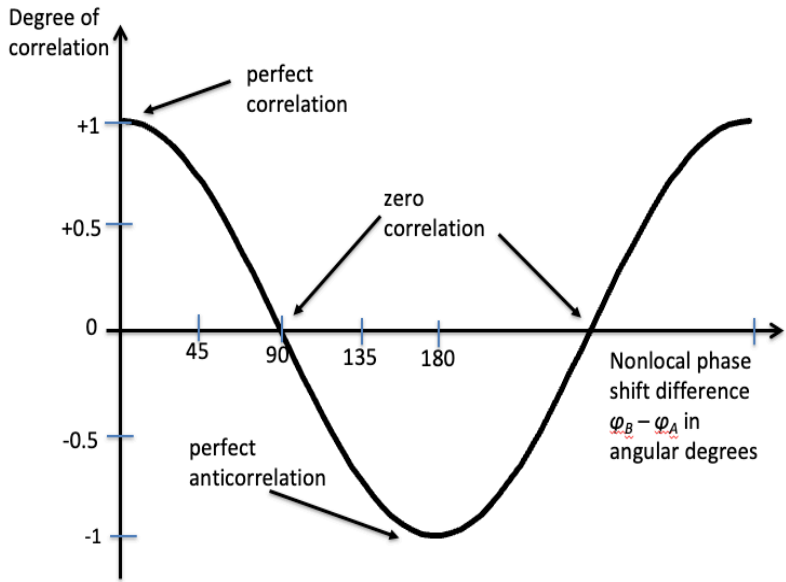

Figure 2.  Nonlocal interference of RTO's bi-photon:  Remarkably, the *degree of correlation* between RTO's two entangled photons varies sinusoidally with the nonlocal phase difference $\phi_B - \phi_A$.

Figure 2 is a remarkable new fact of nature:  When a composite system *AB* with microscopic sub-systems *A* and *B* is superposed (i.e.entangled), *both subsystems* are *rendered incoherent, i.e. the probabilities are phase-independent at all four single-photon detectors:*

$$P(A1) = P(A2) = P(B1) = P(B2) = 0.5 \qquad (8)$$

*regardless of the phase settings $\phi_A$ and $\phi_B$.  While the photons lose their individual phases, the entangled biphoton AB develops a phase*:  The <u>correlation between</u> *A and B is now coherent or "smeared."*

Briefly, *the individual subsystems lose their coherence while the composite system gains coherence.* This "transfer of coherence" from the subsystems to the composite biphoton is the mechanism by which nonlocality at arbitrarily large distances arises in entangled systems.

Indeed, Rarity and Tapster's outcomes at the four detectors violate a Bell inequality, verifying the non-locality. Ou et al. were unable to demonstrate violation of a Bell inequality; they state that "although experiments to demonstrate violations of Bell's inequality" would require higher visibility of the interference, we have nevertheless confirmed the principle of two-photon interference under conditions of very great path difference."

Independently of a violation of Bell's inequality, Figure 2 provides straightforward evidence of this non-locality: Assume the phase shifters satisfy $\phi_A=\phi_B$, (pre-collaboration between the two detection stations would be required to establish this). The two outcomes are then 100% correlated so that *either observer can instantly read off the other observer's outcome simply by glancing at her own detector.* Yet the two stations could be on separate galaxies (see the following alternative set-up).

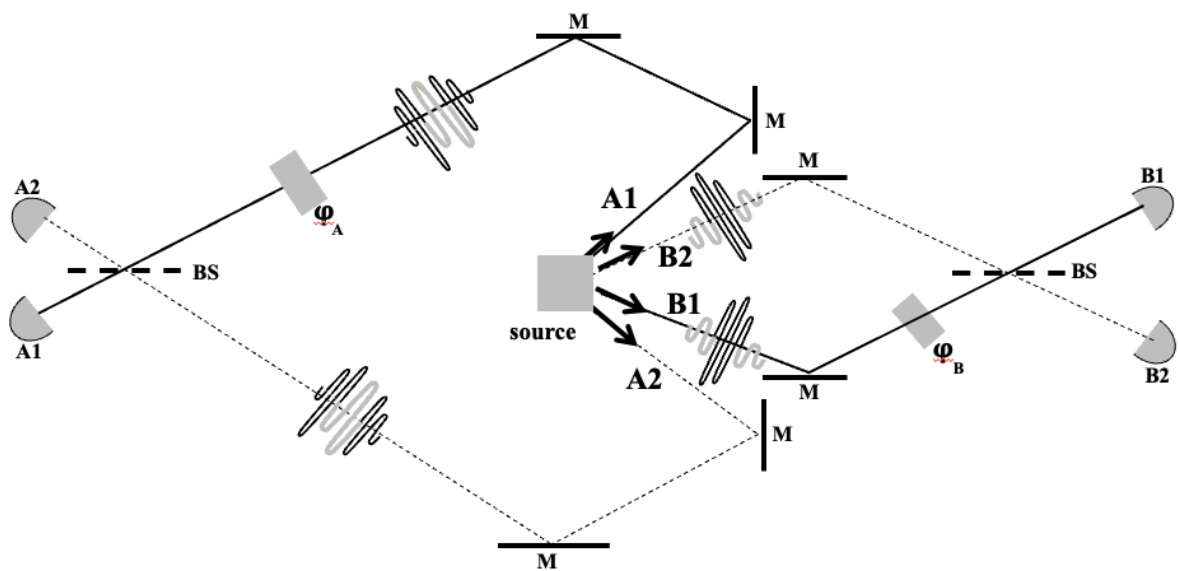

Figure 3. The RTO experiments, designed for widely separated detectors. Compare Figure 1.

If we now return to the pre-measurement state (4), we see that *the measurement problem is solved*: Simply allow subsystem B in (6) to be a macroscopic object such as a Geiger counter. *This* *creates no monstrous macroscopic superposition because B is incoherent and does not go through phases*. Only the *correlation between A and B* goes through phases, as described above. If, for example, *B* is Schrodinger's cat, the cat is not "smeared" (as Schrodinger put it).

The RTO experiments found that, when two quantum objects are entangled, the *correlations between them* become non-local: The degree of interference between *A* and *B* depended upon the difference between their two phases, regardless of their separation distance. *A* and *B* could have been detected in different galaxies yet the non-local effect was instantaneous.

Note that, as the non-local phase (the angle associated with the "plus" sign in the superposition (6)) varies, *the degree of correlation between A and B changes while neither A's state nor B's state changes.* This central point can be summarized as follows: A single quantum system *A* in a simple superposition such as (1) is phase-dependent or "coherent."  *However, when a composite system AB is in a superposition such as the state (6), both subsystems are phase-independent or incoherent while the coherence shifts to the correlations between the two quanta.*

Conclusion: The entangled pre-measurement state (4) is not an absurd macroscopic superposition. Neither subsystem is "smeared" (coherently phase-dependent). Only the *correlations between* the subsystems are coherent, and this is just what we want. The controversial pre-measurement state of a quantum object and its detector is not an absurd superposition of detector states; it is instead a perfectly plausible superposition of correlations between detector states and quantum states of the object. This resolves the measurement problem.

**BRIEF SUMMARY**

The physical meaning of entanglement has been misunderstood. Consider the pre-measurement state:

$$(|A1\rangle|D1\rangle + |A2\rangle|D2\rangle)/\sqrt{2} \qquad (9)$$

where subsystem *A* is a microscopic system such as a single electron passing through a 2-slit experiment, and subsystem *D* could be a pair of which-slit detectors *D1* and *D2* located respectively at the first and second slits, and *|Ai>* and *|Di>* ( i = 1 or 2) are states of *A* and *D*.

The following interpretation of (9) is **INCORRECT**:

Subsystems *A* and *D* are respectively in states *|A1>* and *|D1>* AND subsystems A and *D* are respectively in states |A2> and |D2>     (10)

where "AND" indicates a superposition.  According to (10), the single electron is detected at both slits.  This is absurd and paradoxical.

The following interpretation of this state is **CORRECT**:

"System *A* is in state *|A1>* **IF AND ONLY IF** *D* is in state *|D1>* AND System *A* is in state *|A2>* **IF AND ONLY IF** *D* is in state *|D2>*".     (11)

where "AND" again represents a superposition.  **This is not absurd or paradoxical.**  This resolves the measurement problem.

Comment:  The entangled pre-measurement state (9) is usually described as a "superposition of two product states (or "dyads") *|A1>|D1>* and *|A2>|D2>."* I suggest that (9) instead be described as a "superposition of the *correlations between* two pairs of states." Each dyad *|Ai>|Di>* is a *coherent object*, i.e. an object that goes through phases.  Thus, a simple superposition such as (2) is a superposition of two states of *A,* while an entangled state such as (4) is a *superposition of two correlations between A and D.*  As we have seen, this is not paradoxical.